\documentclass[useAMS,usenatbib,usegraphicx]{mn2e}

\usepackage{color}
\def\lesssim{\mathrel{\hbox{\rlap{\hbox{\lower3pt\hbox{$\sim$}}}\hbox{\raise2pt\hbox{$<$}}}}}
\def\gtrsim{\mathrel{\hbox{\rlap{\hbox{\lower3pt\hbox{$\sim$}}}\hbox{\raise2pt\hbox{$>$}}}}}

\title[An extremely optically dim tidal feature in NGC~871/6/7]{An extremely optically dim tidal feature in the gas-rich interacting galaxy group NGC~871/NGC~876/NGC~877}
\author[K. Lee-Waddell et al.]
	{K.~Lee-Waddell$^{1,2}$\thanks{E-mail: Karen.Lee-Waddell@rmc.ca (KLW)}, 
	K.~Spekkens$^{1}$, J.~-C.~Cuillandre$^{3}$, J.~Cannon$^{4}$, M.~P.~Haynes$^{5}$, 
\newauthor 
	J.~Sick$^{2}$, P.~Chandra$^{6}$, N.~Patra$^{6}$, S.~Stierwalt$^{7}$ and R.~Giovanelli$^{5}$\\
$^{1}$Department of Physics, Royal Military College of Canada, PO Box 17000, Station Forces, Kingston, ON K7K 7B4, Canada\\
$^{2}$Department of Physics, Engineering Physics and Astronomy, Queen's University, Kingston, ON K7L 3N6, Canada\\
$^{3}$Canada-France-Hawaii Telescope Corporation, 65-1238 Mamalahoa Highway, Kamuela, HI, 96743, USA\\
$^{4}$Department of Physics and Astronomy, Macalester College, 1600 Grand Avenue, Saint Paul, MN 55105, USA\\
$^{5}$Center for Radiophysics and Space Research, Space Sciences Building, Cornell University, Ithaca, NY 14853, USA\\
$^{6}$National Centre for Radio Astrophysics, Tata Institute of Fundamental Research, Pune 411 007, India\\
$^{7}$Department of Astronomy, University of Virginia, 530 McCormick Road, Charlottesville, VA 22904, USA}

\begin{document}

\date{Accepted 2014 July 3. Received 2014 July 1; in original form 2014 March 24}

\pagerange{\pageref{firstpage}--\pageref{lastpage}} \pubyear{2014}

\maketitle

\label{firstpage}

\begin{abstract}

\noindent We present GMRT {H\sc{i}} observations and deep CFHT MegaCam optical images of the gas-rich interacting galaxy group NGC~871/NGC~876/NGC~877 (hereafter NGC~871/6/7).  Our high-resolution data sets provide a census of the {H\sc{i}} and stellar properties of the detected gas-rich group members.  In addition to a handful of spiral, irregular and dwarf galaxies, this group harbours an intriguing {H\sc{i}} feature, AGC~749170, that has a gas mass of $\sim$$10^{9.3}M_\odot$, a dynamical-to-gas mass ratio of $\sim$1 (assuming the cloud is rotating and in dynamical equilibrium) and no optical counterpart in previous imaging.  Our observations have revealed a faint feature in the CFHT $g'$- and $r'$-bands; if it is physically associated with AGC~749170, the latter has $M/L_g >1000 M_{\odot}/L_{\odot}$ as well as a higher metallicity (estimated using photometric colours) and a significantly younger stellar population than the other low-mass gas-rich group members.  These properties, as well as its spectral and spatial location with respect to its suspected parent galaxies, strongly indicate a tidal origin for AGC~749170.  Overall, the {H\sc{i}} properties of AGC~749170 resemble those of other optically dark/dim clouds that have been found in groups.  These clouds could represent a class of relatively long-lived {H\sc{i}}-rich tidal remnants that survive in intermediate-density environments.

\end{abstract}

\begin{keywords}
galaxies: interactions -- galaxies: dwarf
\end{keywords}


At the current epoch, most galaxies can be found in medium density group environments (e.g. \citealt{e2004}, \citealt{t2008}), where tidal interactions within these systems play important roles in galactic evolution.  Several nearby, gas-rich groups exhibit clear signs of these interactions (e.g. \citealt{f2009}), providing a snapshot of this evolutionary process in action.  Material pulled from spiral-rich interacting systems forms gaseous filaments that can harbour second-generation tidal knots and tidal dwarf galaxies (TDGs), which differ from classical satellite dwarfs -- dwarf galaxies have masses $\lesssim 10^9 M_{\odot}$ -- by their lack of dark matter, higher metallicity content and higher star formation rates (e.g. \citealt{b2004}).   

In recent neutral hydrogen ({H\sc{i}}) surveys of groups and clusters, large detached gaseous clouds with no apparent stellar component have also been detected (e.g. \citealt{k2000}, \citealt{r2001},  \citealt{k2007}).  Some of these objects were initially suspected as being extremely rare primordial ``dark galaxies'' and gained considerable amounts of attention (e.g. \citealt{m2005}); however, due to their environment, it is currently believed that these so-called dark galaxies are actually another form of tidal debris (e.g. \citealt{b2005}, \citealt{h2007}, \citealt{d2008}).  The number of detected optically dark clouds is growing and although most of the galaxy-sized {H\sc{i}}-rich clouds are found in regions where interaction events and other environmental processes could explain their origin, their actual formation properties are still unclear  (e.g. \citealt{w2005}, \citealt{e2010}, \citealt{k2010}, \citealt{m2012}, \citealt{o2013}, \citealt{s2013}).  Broad searches for isolated {H\sc{i}} clouds have yet to confirm the existence of primordial dark galaxies (e.g. \citealt{d2005}, \citealt{w2009}, \citealt{h2011}). 

In concert with the distributions of intra-group gas and stars, tidal and classical dwarf galaxy populations provide important evolutionary diagnostics and cosmological constraints in group environments.  For merging systems, the formation of any tidal object also greatly constrains the type of interaction and the properties of the original objects involved in the process (see \citealt{k2012}).  Although TDGs are frequently produced in simulations and several candidates have been reported (see \citealt{sh2009} and references therein), very few are widely accepted as authentic.  One of the main reasons for this discrepancy is the variety of corroborating observations that are required to unambiguously distinguish between classical satellites, TDGs and unbound tidal knots (e.g. \citealt{d2011}).  

Given the properties of tidal systems as well as the environments in which they form, {H\sc{i}} observations are useful preliminary search tools as they can indicate regions of potential star formation, trace the location of tidally formed features and have been routinely used to map the gas distribution in and around various systems (e.g. \citealt{k2009}, \citealt{st2009}).  The Arecibo Legacy Fast ALFA survey (ALFALFA; \citealt{g2005}) unbiasedly maps {H\sc{i}} in the local volume.  The recent data release of 40 per cent of the ALFALFA catalog ($\alpha.40$; \citealt{h2011}) has a wealth of {H\sc{i}}-rich objects with intriguing features that warrant further follow-up.  In this paper, we target NGC~871/NGC~876/NGC~877 (hereafter referred to as NGC~871/6/7), a galaxy group with a fairly extensive {H\sc{i}} envelope.
     
As shown in \citeauthor{l2012} (2012; hereafter referred to as Paper 1), high-resolution {H\sc{i}} imaging using the Giant Metrewave Radio Telescope (GMRT) is able to resolve the sizes and dynamics of low-mass gas-rich galaxies on kpc scales.  The configuration of the array can produce {H\sc{i}} images with various synthesized spatial resolutions, from a single set of data, using post-observing processing techniques (see Section 1.1).  As well, the versatility of the recently added GMRT software correlator can produce high spectral resolution observations (see \citealt{r2010} for details on the software processing pipeline).

The properties of the stellar populations in dwarf galaxies also help distinguish tidal systems.  The Canada-France-Hawaii Telescope (CFHT) MegaCam used in conjunction with the Elixir+Elixir-LSB processing pipeline, which has been developed for the deep optical surveys used in the ATLAS$^{\mbox{3D}}$ and NGVS projects (\citealt{d2011}, \citealt{f2012}), can produce high resolution optical images down to a limit of 29 mag arcsec$^{-2}$ in the $g'$-band (Cuillandre et al., in preparation).  This technique is a powerful tool for constraining the evolutionary histories of low surface brightness detections in the local volume.  For example, \citet{m2010} use Elixir+Elixir-LSB to reveal optical counterparts to several {H\sc{i}} condensations along the Leo Ring, in the $g'r'i'$-bands, suggesting a collisional origin for this feature.  Similarly, a deep analysis of the stellar components of the NGC~871/6/7 group members will give insight into their properties and can eventually be used constrain the interaction history of the system (see \citealt{p2013}).  

\begin{figure}
  \includegraphics[width=84mm]{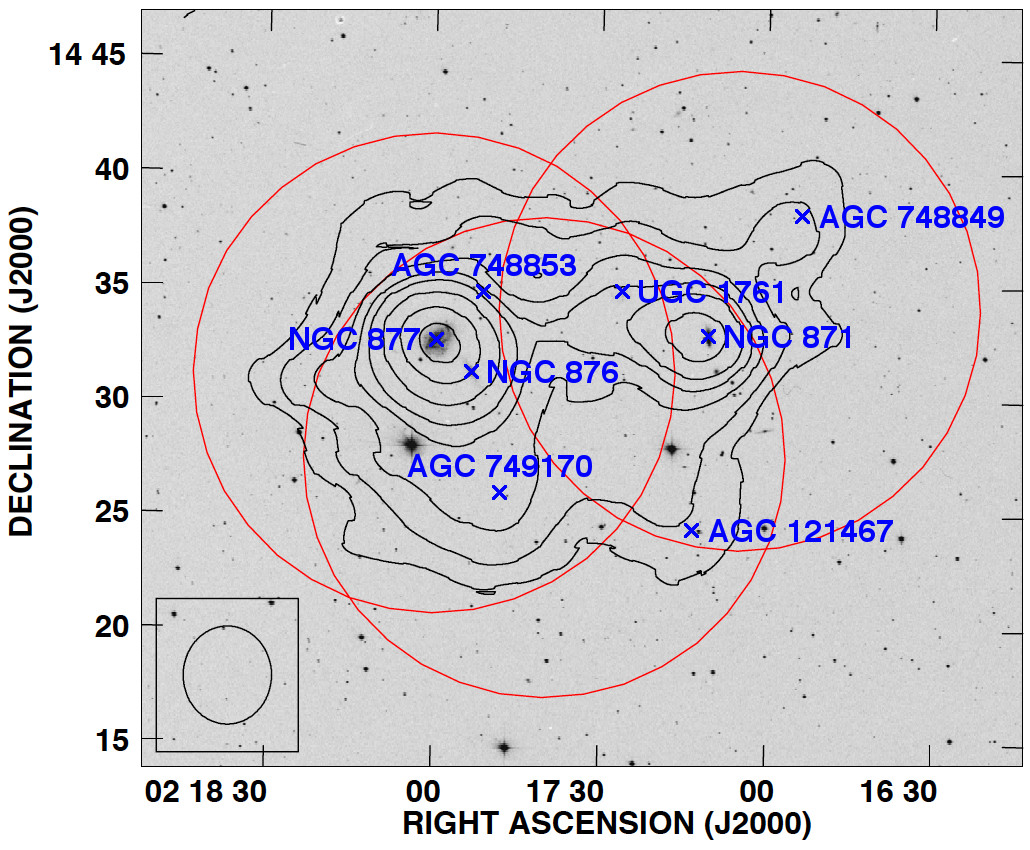}
  \caption{ALFALFA total {H\sc{i}} intensity contours at $N_{H_I} = (1.8, 4, 8, 12, 16, 24, 36, 48) \times 10^{19}$ atoms cm$^{-2}$ superimposed on a DSS2 \textit{r}-band greyscale image of the NGC~871/6/7 group.  The heliocentric velocity of the {H\sc{i}} contained within the lowest contour ranges from 3550 to 4000 km s$^{-1}$.  The 4 arcmin (60 kpc) ALFALFA beam is in the bottom left corner and the blue x's indicate the locations of the ALFALFA {H\sc{i}} detections.  The large red circles show the GMRT follow-up pointings. \label{ALFA}}
\end{figure}

NGC~871/6/7 is a gas-rich interacting group located at a distance of $\sim$50 Mpc and consists of at least three spirals: NGC~871, NGC~876 and NGC~877 and one irregular galaxy: UGC~1761 (\citealt{sp2007}, \citealt{sp2009}; properties listed in Table \ref{NGC}).  This group resides in a common {H\sc{i}} distribution, as mapped by ALFALFA, that spans $0.5 \deg^2$ and has a total {H\sc{i}} mass of $M_{H_I} > 6 \times 10^{10} M_{\odot}$ (Fig.~\ref{ALFA}; \citealt{h2011}), which is unusually high for a group of this size and composition: galaxies with $M_{H_I} > 10^{10} M_{\odot}$ are rare in the local universe (e.g. \citealt{mar2010}) and according to $\alpha$.40, the {H\sc{i}} mass of each large spiral in the group exceeds this value.  Few ($<$2 per cent) of the galaxies with $M_{H_I} > 10^{10} M_{\odot}$ in $\alpha$.40 are as tightly clustered in triplets on the sky and in velocity as NGC~871/6/7, which is the closest (in terms of distance from the sun) example of such systems \citep{h2011}.

Within the spatial and spectral location of the group, $\alpha$.40 also indicates four low-mass gas-rich objects with $M_{H_I}\sim10^9 M_{\odot}$ (see Section 2).  One of these {H\sc{i}} peaks, AGC~749170, appeared to have no detectable stellar component in the existing data and warranted detailed follow-up.  ``First-look'' observations using the Karl G. Jansky Very Large Array (VLA) in C-configuration completed in 2009, from program AC963 and PI J. Cannon, confirmed the spatial and spectral location of the eight gas-rich group members. 

In this paper, we present our multi-wavelength follow-up campaign of NGC~871/6/7.  Using high-resolution observations from the GMRT, we have isolated the {H\sc{i}} belonging to each gas-rich group member, thereby enabling gas mass calculations and dynamical mass estimates (Section 2).  Deep optical imaging from the CFHT MegaCam and ancillary Galaxy Evolution Explorer (\textit{GALEX}) data were used to identify likely optical counterparts as well as estimate the stellar properties for detectable low {H\sc{i}} mass -- $M_{H_I} \lesssim 10^{9.2} M_{\odot}$ -- group members (Section 3).  We then discuss our results for AGC~749170 and its likely tidal origin while comparing its properties to similar objects found in the literature (Section 4).

\begin{table}
\centering
 \begin{minipage}{85mm}
\caption{General properties of the large galaxies within NGC~871/6/7.  Column 1: common name; column 2: right ascension and declination; column 3: galaxy morphology (RC3.9; \citealt{c1994}); column 4: heliocentric velocity \citep{sp2005}}
\label{NGC}
\begin{tabular}{@{}l c c c c@{}} 
\hline
Name		&Coordinates				&Morphology	&$cz_{\odot}$	\\
			&(J2000)					&				&(km s$^{-1}$) 	\\
(1)			&(2)						&(3)			&(4)			\\
\hline 
NGC~871		&02 17 11, +14 32 53	&SB(s)c			&3740 $\pm$ 1	\\
UGC~1761		&02 17 26, +14 34 49 	&Im				&4010 $\pm$ 5	\\
NGC~876		&02 17 53, +14 31 16 	&SAc 			&3894 $\pm$ 1	\\ 
NGC~877		&02 18 00, +14 32 39 	&SAB(rs)bc		&3913 $\pm$ 1	\\
\hline 
\end{tabular}
\end{minipage}
\end{table}

\section{Observations of NGC~871/6/7}

\subsection{GMRT Observations}

Using the ALFALFA map as a guide and following a similar observational set-up and reduction technique as Paper 1, the GMRT data consist of three pointings observed over a total of four nights in the late summer of 2012 (Fig.~\ref{ALFA}).  Due to technical difficulties with the array, the observations from the first night were unusable and an additional eight hours of observing were granted.  The 24h of usable telescope time, which included calibration observations using standard flux calibrators (3C48 and 3C147) and a nearby phase calibrator 0204+152, were equally divided between the three pointings.  The observing and map parameters are summarized in Table \ref{param}.

\begin{table}
 \centering
 \begin{minipage}{85mm}
 \caption{GMRT observation set-up and image properties.}
 \label{param}
\begin{tabular}{@{} l c c@{}}  
\hline
Parameter 								& Value		& Units		\\ 
\hline 
Number of pointings						& 3			&			\\
Average time on source per pointing		& 320		& min		\\ 
Primary beam HPBW per pointing			& 19.7		& arcmin	\\
Mosaicked map size						& 30		& arcmin	\\ 
Central observing frequency				& 1402.2	& MHz		\\
Observing bandwidth						& 4.16		& MHz		\\ 
Observing spectral resolution				& 8.1		& kHz		\\ 
Final cube spectral resolution				& 24.4		& kHz		\\
Final cube spectral resolution				& 5.3		& km s$^{-1}$\\
Map spatial resolutions					& 30 and 45	& arcsec	\\ 
Peak map sensitivity (45 arcsec res.)		& 1.6	& mJy beam$^{-1}$\\  
Peak map sensitivity (30 arcsec res.)		& 1.3	& mJy beam$^{-1}$\\  
\hline
\end{tabular} 
\end{minipage}
\end{table}

\begin{figure*}
\begin{center}
  \includegraphics[width=168mm]{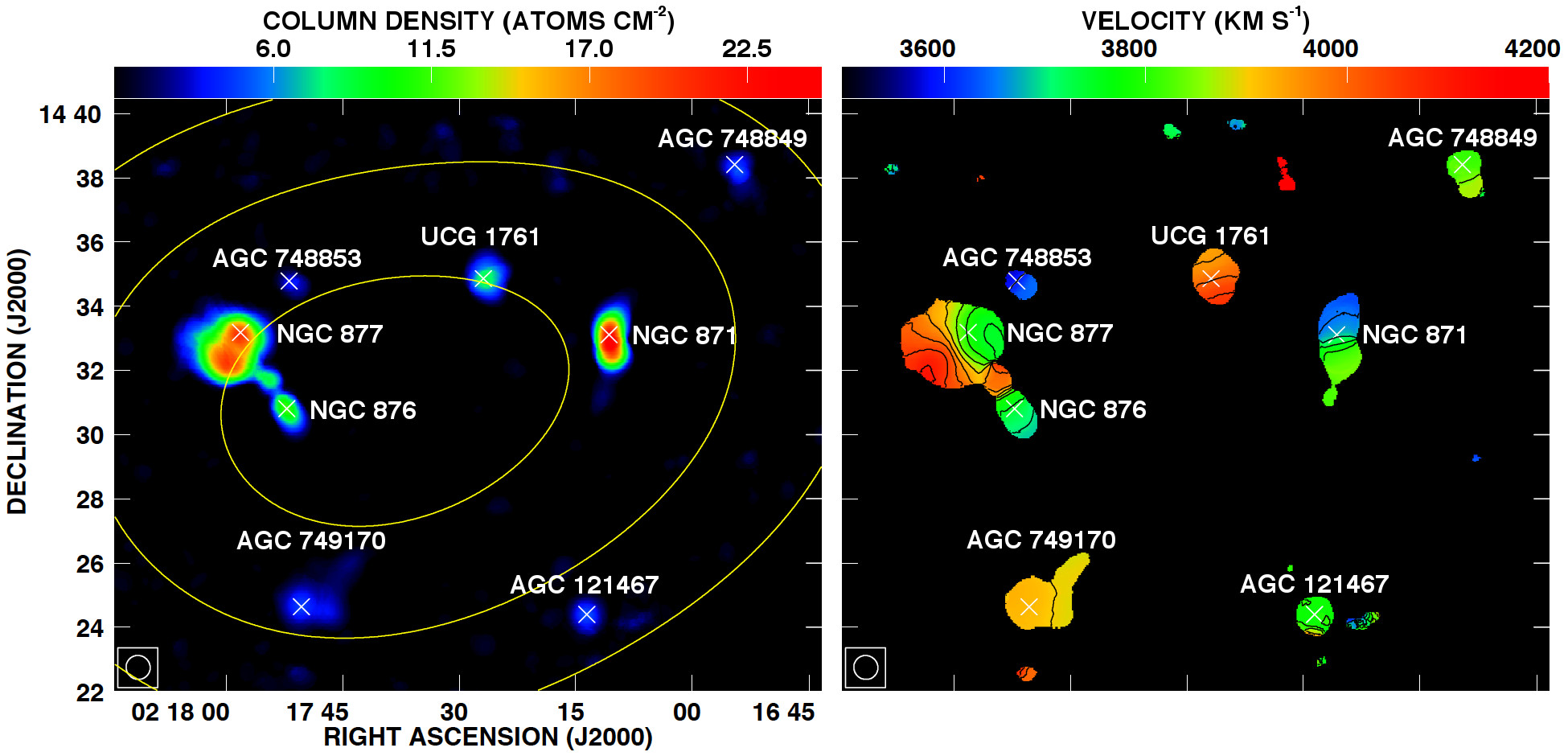}
  \caption{GMRT 45 arcsec angular resolution maps of the {H\sc{i}} in NGC~871/6/7.  The x's indicate the peak {H\sc{i}} flux density of all gas-rich detections (see Table~\ref{HI_mass}, column 2) and the synthesized beam is shown in the bottom left corner of each panel.  Left: Total intensity map (mom$_0$).  The yellow contours show the regions in the map with map noise $\sigma = (1.7, 2, 2.5)$ mJy beam$^{-1}$ and the colour scale ranges linearly from 1 to 25 $\times10^{20}$ atoms cm$^{-2}$.  Right: Intensity-weighted velocity map (mom$_1$).  Velocity contours are shown at 50 km s$^{-1}$ increments.  For both maps, in order to remove spurious signals, regions in each datacube channel where $\sigma > 2.5$ mJy beam$^{-1}$ were blanked before the moments were computed.  In addition, the velocity map is blanked at locations where the column density ($N_{H_I}$) $\leq 1\times10^{20}$ atoms cm$^{-2}$. \label{GMRT}}
\end{center}
\end{figure*}

Data editing and reduction was completed using the Astronomical Image Processing System (AIPS) version 31Dec12 \citep{g2003} in the same manner as presented in Paper 1.  The calibrated data cubes were mosaicked and imaged with natural weighting and tapering to respectively produce $\sim$45 arcsec (11 kpc at the distance of NGC~871/6/7) and $\sim$30 arcsec (7 kpc) synthesized beams, which constrain the {H\sc{i}} mass, size and structure of the low-mass features while minimizing the noise.  A three-channel spectral average, resulting in a resolution of 5.3 km s$^{-1}$ (complementary to the ALFALFA resolution of 5.2 km s$^{-1}$) was used to produce the final image maps as shown in Fig. \ref{GMRT}, which indicates eight gas-rich objects.

\subsection{CFHT Observations}

\begin{figure*}
\begin{center}
  \includegraphics[width=168mm]{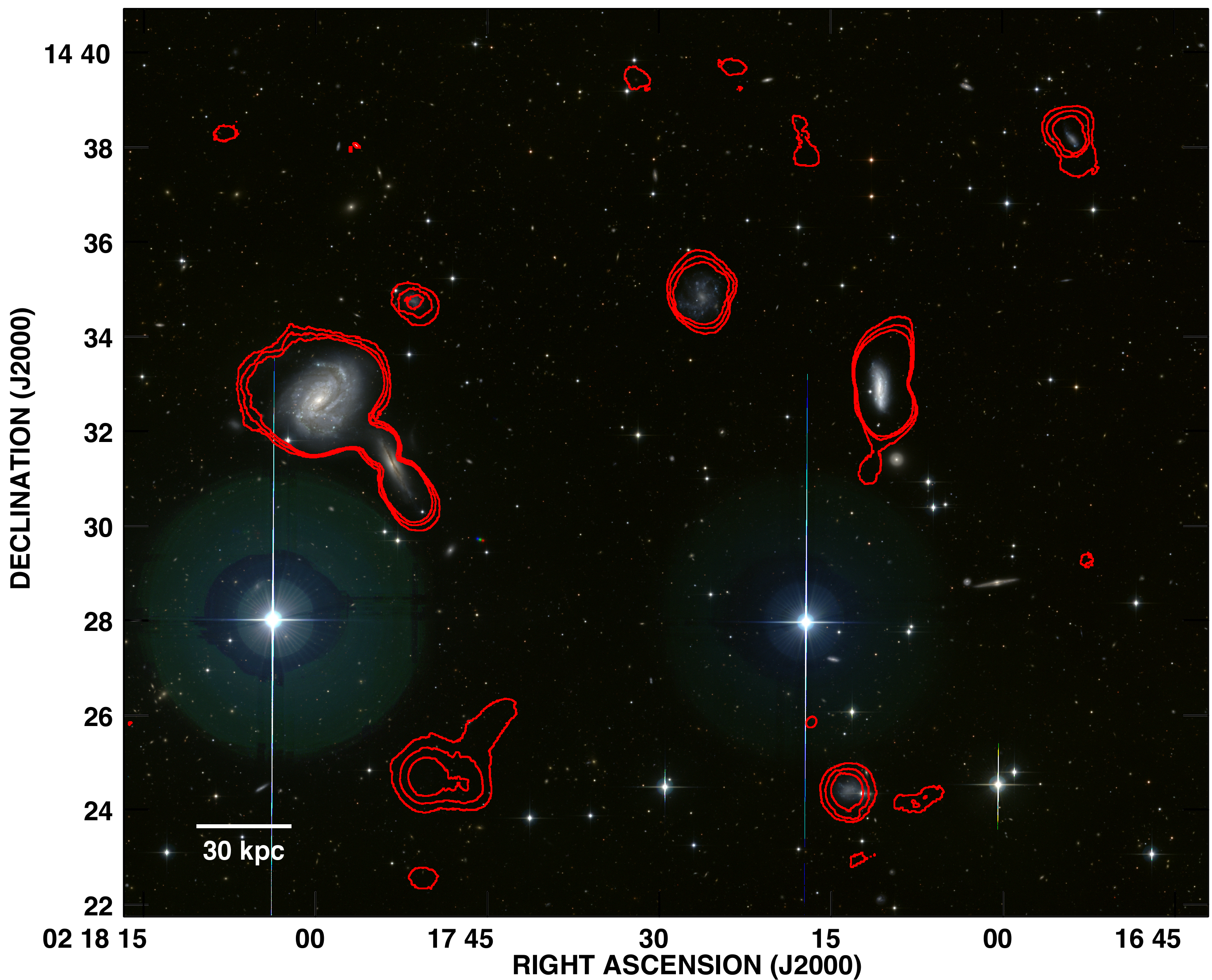}
  \caption{CFHT $g'r'i'$-band composite image with the GMRT 45 arcsec resolution intensity contour at $N_{H_I} = (1, 2, 3) \times 10^{20}$ atoms cm$^{-2}$ to show the span of {H\sc{i}} associated with each gas-rich detection. 
  \label{CFHT}}
\end{center}  
\end{figure*}

Using the CFHT MegaCam, we obtained $g'r'i'$-band imaging of NGC~871/6/7 in September 2012.  MegaCam's 1~deg$^2$ field of view imaged the $0.5 \deg^2$ group while simultaneously observing sufficient background to allow for sky modelling and subtraction using the Elixir+Elixir-LSB processing pipeline.  Each filter used a seven-point large dithering pattern (LDP-CCD-7) with exposures sequenced within a one hour time window in order to minimize sky variations.  The MegaCam images were stacked using Elixir to characterize and subtract the background and then processed through Elixir-LSB to remove the scattered light components \citep{f2012}.  The final image pixels are binned $3\times3$ to boost the signal-to-noise ratio of the background.  The resultant images are at least five times more sensitive than the SDSS optical imaging limit \citep{s2002}. The set-up parameters and resulting image properties are summarized in Table \ref{MegaCam}.  A composite image is shown in Fig.~\ref{CFHT} along with our GMRT {H\sc{i}} detections.  

\begin{table}
 \centering
 \begin{minipage}{85mm}
 \caption{CFHT MegaCam observation set-up and image properties.}
 \label{MegaCam}
\begin{tabular}{@{} l @{} c c c@{}}  
\hline
Parameter 							& $g'$-band		& $r'$-band		&$i'$-band	\\ 
\hline 
Exposure time (sec)					& 280  			& 280			&194		\\ 
Number of exposures				& 7				& 7				& 7			\\
Mean image quality (arcsec)			& 0.71			& 0.64			& 0.59		\\
Sky background, $3\times3$ bin (ADU)				&387.8			& 942.9			&1699.1	\\
$3\times3$ sky brightness (mag arcsec$^{-2}$)	& 21.9			& 20.9			& 20.3		\\
$3\times3$ detection limit  (mag arcsec$^{-2}$)	& 27.3			& 26.6			& 26.0		\\ 
\hline
\end{tabular} 
\end{minipage}
\end{table}

Subtraction of contamination from foreground stars and background galaxies used the \citet{s1995} ring-filter technique embedded within Elixir+Elixir-LSB.  The $\sim$6 arcmin (90 kpc) diameter reflection halos in Fig.~\ref{CFHT} from two bright stars, HD 14108 (V = 8.62 $\pm$ 0.02 mag) and HD 14192 (V = 7.67 $\pm$ 0.01 mag), within the field of view at 02 17 17.1, +14 27 57 and 02 18 03.7, +14 28 00 could not be excised and have left artifacts in the final optical images.  Fortunately, these artifacts lie just outside the spatial location of our gas-rich detections.


\section{{H\sc{i}} Data Analysis}

The eight gas-rich group members (three large spirals, one large irregular and four low-mass features) detected by ALFALFA were also detected in the GMRT data as shown in Fig. \ref{GMRT}.  No additional {H\sc{i}} sources were detected in the spatial and spectral region of the group.  The {H\sc{i}} column density is computed using:
\begin{equation}
N_{H_I} \mbox{[atoms cm$^{-2}$]} = \frac{2.228 \times 10^{24}}{\theta_1 \theta_2 \nu_c^2} \int I_{\nu} dv
\end{equation}
where $\theta_1$ and $\theta_2$ are the semi-major and semi-minor axes of the beam in arcsec, $\nu_c$ is the central band frequency in GHz and $\int I_{\nu} dv$ is the contour level in Jy beam$^{-1}$ km s$^{-1}$ from the mom$_0$ maps \citep{i2009}.  Equation 1 assumes that the source uniformly fills the beam and therefore corresponds to an average column density per resolution element.   

\begin{figure*}
\begin{center}
  \includegraphics[width=84mm]{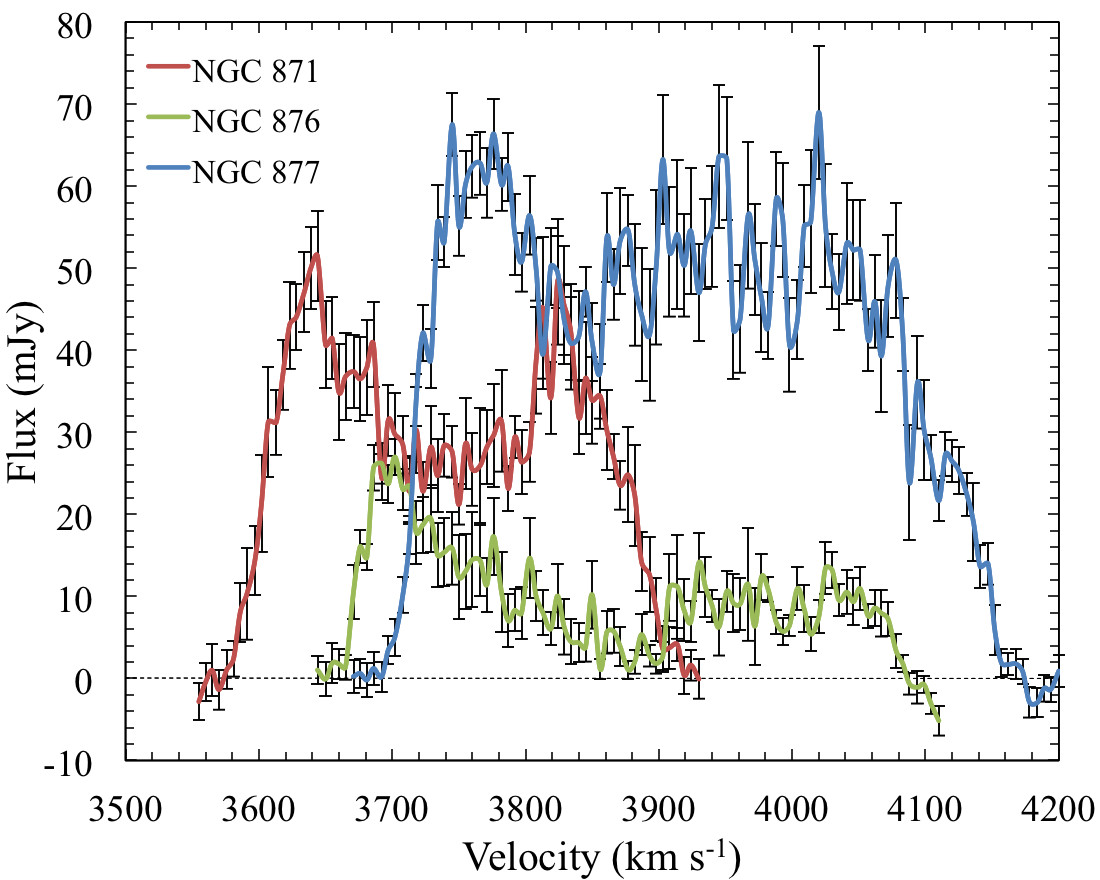}
    \includegraphics[width=84mm]{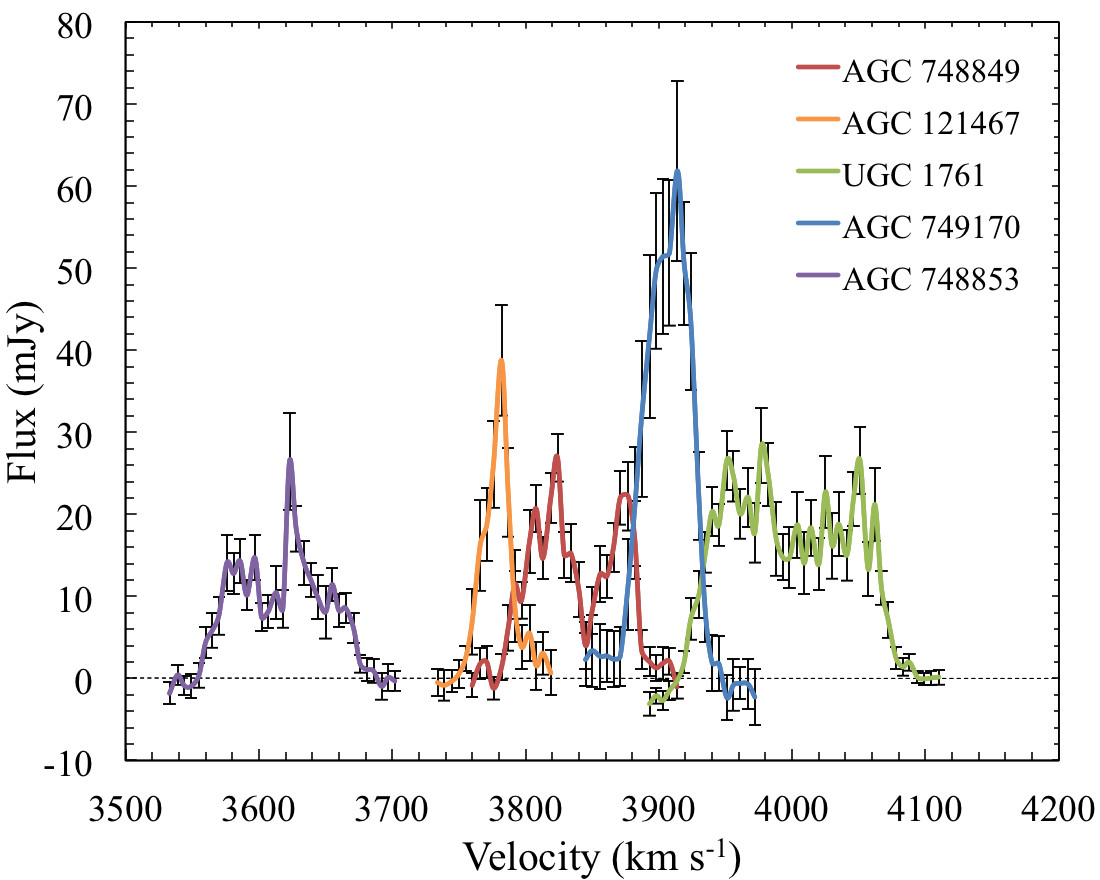}
  \caption{Global profiles derived from the GMRT 45 arcsec resolution data.  Left: Large spiral galaxies.  Right: Low {H\sc{i}} mass group members. 
  \label{GP}}
\end{center}
\end{figure*}

Measurements of the flux density from the higher sensitivity 45 arcsec resolution GMRT maps were used to produce the global profiles in Fig.~\ref{GP}.  As in Paper 1, the {H\sc{i}} mass is calculated using:
\begin{equation}
M_{H_I} [M_{\odot}] = 2.356 \times 10^5 d^2 S_{H_I}
\end{equation}
where $d$ is the distance to the source in Mpc and $S_{H_I}$ is the flux density of the global profiles integrated over velocity.  The basic properties of the GMRT {H\sc{i}} detections are in Table \ref{HI_mass}.  A comparison of the GMRT data (which are consistent with the results from the noisier first-look VLA data) and ALFALFA observations indicate that a significant portion of the gas ($\sim$50 per cent) is diffusely distributed around the interacting spirals and throughout the system on scales greater than $\sim$5 arcmin (70 kpc).

\begin{table*}
\centering
 \begin{minipage}{168mm}
\caption{{H\sc{i}} detections in NGC~871/6/7.  Column 1: detection name; column 2: right ascension and declination of peak {H\sc{i}} flux density; column 3: centroid of the most probable optical counterpart.  An optical feature coincident with AGC~749170 has been detected, for the first time, during this work; column 4: noise in region of detection in 45 arcsec GMRT maps; column 5: integrated {H\sc{i}} flux density computed from the global profiles in Fig.~\ref{GP}.  The uncertainty was determined by propagating the error from the global profile and adding a 10 per cent calibration error; column 6: {H\sc{i}} mass calculated from GMRT observations; column 7: {H\sc{i}} mass from ALFALFA $\alpha.40$ data release \citep{h2011}}
\label{HI_mass}
\begin{tabular}{@{}l c c c c c c@{}} 
\hline
Name	&GMRT Coordinates		&Optical Coordinates		&$\sigma$			&$S_{H_I}$			&$M_{H_I}$(GMRT)		&$M_{H_I} (\alpha.40)$\\
		&(J2000) 				&(J2000)				&(mJy beam$^{-1}$)	&(Jy km s$^{-1}$)	&($10^8 M_{\odot}$)		&($10^8 M_{\odot}$)\\
(1)		&(2) 					& (3) 					& (4)				&(5)					&(6)						&(7)\\	
\hline
AGC~748849	&02 16 54.5, +14 38 25	&02 16 53.8, +14 38 09	& 2.3		&1.5 $\pm$ 0.4 		& 9 $\pm$ 2 				&14.4 $\pm$ 0.4\\
NGC~871		&02 17 10.7, +14 33 07 	&02 17 10.5, +14 32 53	& 1.8		& 10 $\pm$ 2 		& 60 $\pm$ 10 			&112.8 $\pm$ 0.7\\
AGC~121467	&02 17 13.6, +14 24 24	&02 17 13.4, +14 24 24  	& 2.2		& 0.8 $\pm$ 0.2 		& 5 $\pm$ 1				&14.7 $\pm$ 0.3\\ 
UGC~1761		&02 17 27.4, +14 34 52	&02 17 26.3, +14 34 49 	& 1.7		& 2.8 $\pm$ 0.7 		& 16 $\pm$ 4			&29.6 $\pm$ 0.4\\ 
AGC~749170 	&02 17 50.3, +14 24 40	&02 17 50.2, +14 24 40	& 1.9		& 2.4 $\pm$ 0.6 		& 14 $\pm$ 4 			&29.9 $\pm$ 0.3\\
AGC~748853	&02 17 51.9, +14 34 47	&02 17 51.4, +14 34 45 	& 1.8		& 1.3 $\pm$ 0.3 		& 8 $\pm$ 2 				&11.6 $\pm$ 0.4\\
NGC~876		&02 17 52.2, +14 30 49 	&02 17 53.2, +14 31 16	& 1.6		& 4 $\pm$ 1 			& 26 $\pm$ 6 			&162.0 $\pm$ 0.8\\	
NGC~877		&02 17 58.2, +14 33 11	&02 17 59.6, +14 32 39	& 1.7		& 21 $\pm$ 4 		& 120 $\pm$ 30 			&219.3 $\pm$ 0.9\\
\hline 
\end{tabular}
\end{minipage} 
\end{table*}

Assuming that each gas-rich group member is self-gravitating and in dynamical equilibrium, we estimate the total dynamical mass of their {H\sc{i}} regions using:
\begin{equation}
M_{dyn} [M_{\odot}] = 3.39 \times 10^4 a_{H_I} d \left( \frac{W_{20}}{2\sin i} \right)^2
\end{equation}
where $a_{H_I}$ is the {H\sc{i}} major axis diameter of the object in arcmin, $d$ is the distance to the source in Mpc and $W_{20}$ is the velocity width, at 20 per cent of the peak flux density, of each object in km s$^{-1}$.  The inclination ($i$) for the NGC galaxies is derived from the objects' axis ratios ($a/b$ where $a$ is the major axis and $b$ is the minor axis) as recorded in the HyperLeda database \citep{p2003}; whereas, $a/b$ calculated by SExtractor (\citealt{b1996}; see Section 3) is used for UGC~1761 and the low-mass features where
\begin{equation}
\cos^2 i = \frac{(b/a)^2 - q_o^2}{1-q_o^2}
\end{equation}
with the intrinsic flatness of a galaxy assumed to be $q_o = 0.20$ \citep{h1972}.  

\begin{figure*}
  \includegraphics[width=168mm]{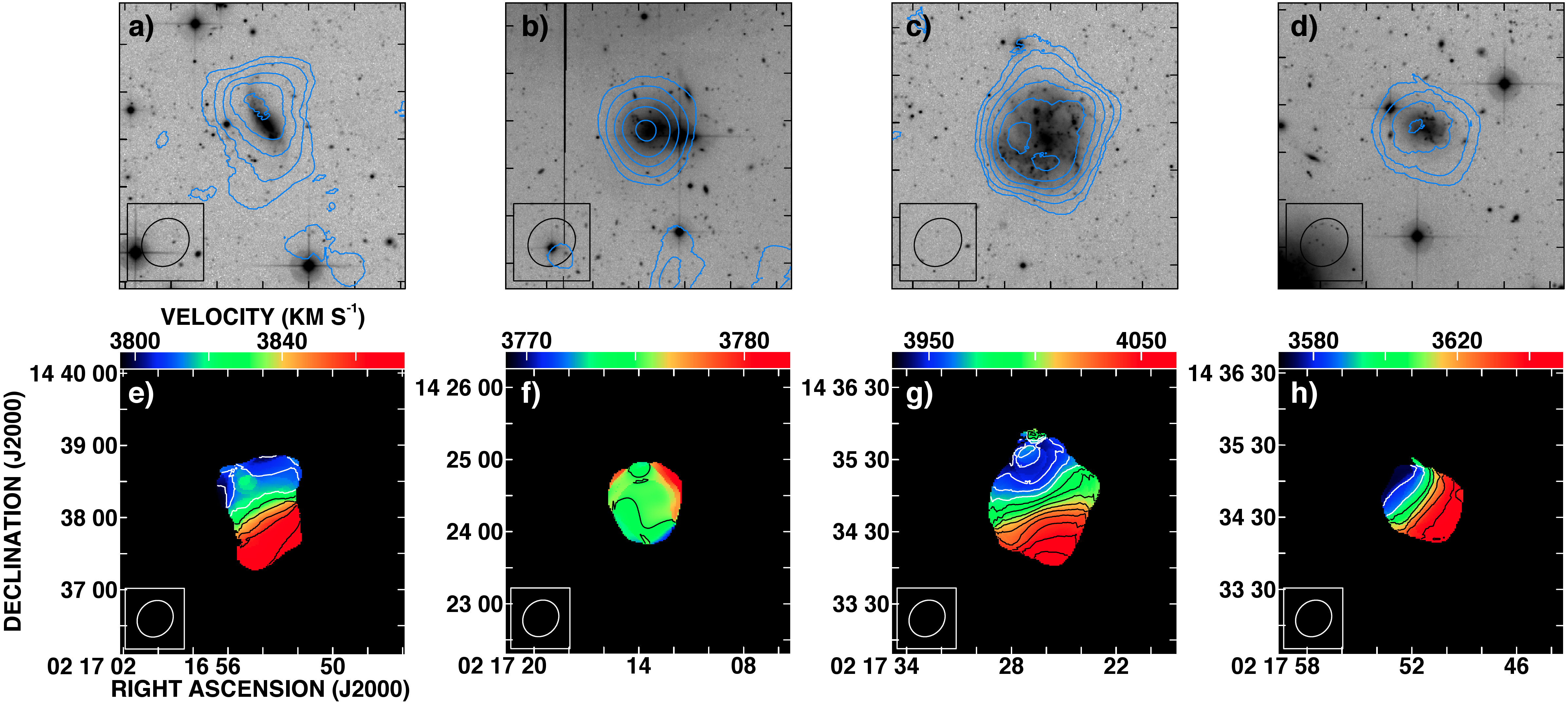}
  \caption{Top: GMRT mom$_0$ maps at 30 arcsec resolution superimposed on CFHT \textit{g'}-band images.  Contours are at $N_{H_I} = (1, 2, 3.5, 5, 7, 9.5) \times 10^{20}$ atoms cm$^{-2}$.  Bottom: GMRT mom$_1$ maps at 30 arcsec resolution.  The velocity contours are at increments of 10 km s$^{-1}$.  (a and e) AGC~748849; (b and f) AGC~121467; (c and g) UGC~1761; (d and h) AGC~748853. \label{dwarfs}}
\end{figure*}

\begin{figure*}
  \includegraphics[width=168mm]{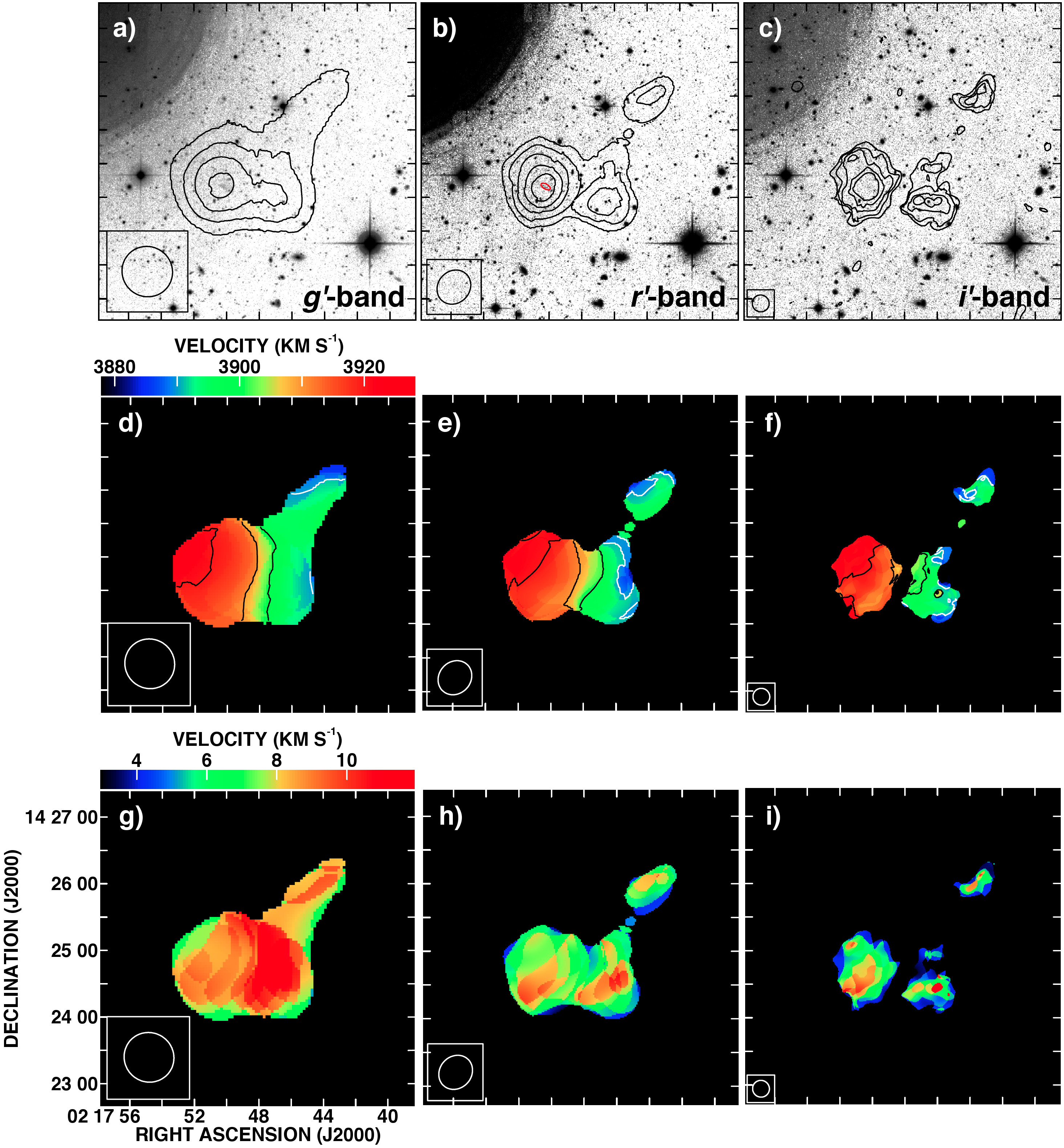}
  \caption{Top: GMRT mom$_0$ maps at various angular resolutions (45 arcsec for a, d and g; 30 arcsec for b, e and h; 15 arcsec for c, f and i) of AGC~749170 superimposed on three bands of CFHT images.  The black contours are at (a) $N_{H_I} = (1, 2, 3, 4) \times 10^{20}$ atoms cm$^{-2}$ (b) $N_{H_I} = (1, 2, 3, 4, 5) \times 10^{20}$ atoms cm$^{-2}$ (c) $N_{H_I} = (1, 3, 5, 7, 9) \times 10^{20}$ atoms cm$^{-2}$.  The red ellipse in (b) indicates the region used by SExtractor to calculate the optical properties (see Section 3). Middle: GMRT mom$_1$ maps.  Velocity contours are at increments of 10 km s$^{-1}$.  Bottom: GMRT second velocity moment maps (mom$_2$).  \label{dark}}
\end{figure*}

For each GMRT detection, $a_{H_I}$ was measured from the 30 arcsec angular resolution maps (e.g. Figs \ref{dwarfs} and \ref{dark}e) perpendicular to the axis of rotation, out to a column density of $1 \times 10^{20}$ atoms cm$^{-2}$, and corrected for beam smearing by deconvolving the lengths with a Gaussian representing the beam size.  The dynamical masses of each object computed using the velocity width of the integrated spectra (i.e. $W_{20}$) are, on average, 40 per cent higher than the inclination corrected total masses estimated from the maximum velocity difference across $a_{H_I}$ (i.e. the difference in the velocity between the extreme edges of each object).  In order to produce more conservative estimates of the dynamical to baryonic mass fraction for each low {H\sc{i}} mass feature (see Section 3), the estimates of the dynamical mass computed using $W_{20}$ are presented in Table~\ref{dyn} with uncertainties of half the difference between the masses computed using the two methods.  

Note it was not possible to extract a reliable inclination for AGC~121467 because of an overlapping foreground star (see Fig. \ref{dwarfs}b).  This object also shows no distinguishable velocity gradient perpendicular to the line of sight in Fig. \ref{dwarfs}f.  Accordingly, the dynamical mass for AGC~121467 has not been computed.  The assumption of dynamical equilibrium is likely valid for AGC~748849, UGC~1761 and AGC~748853, given the coincidence of the {H\sc{i}} and optical counterparts, the clear velocity gradient across the {H\sc{i}} detection and the optical morphology of the counterparts (Fig. \ref{dwarfs}).

Fig. \ref{dark} shows the total intensity and various velocity maps of AGC~749170 at three different angular resolutions.  Additional tapering was applied to the data to produce the 15 arcsec resolution maps of AGC~749170 in an attempt to further investigate its structure.  The datacube at this level of spatial resolution is quite noisy and accordingly, corresponding {H\sc{i}} images were not produced for the other group members.  The top row of images in Fig. \ref{dark} reveal, for the first time, a possible faint optical counterpart -- located at the point of highest $N_{H_I}$ -- in the $g'$ and $r'$-bands for this previously dark feature.  Nevertheless, the feature is not visible in the noisier $i'$-band data.  The first and second velocity maps for AGC~749170 (Fig. \ref{dark} middle and bottom rows) show a velocity gradient.  We discuss whether or not this gradient represents a self-gravitating, rotating disk in Section 4.

\begin{table*}
\centering
 \begin{minipage}{93mm}
\caption{Dynamical information for NGC~871/6/7 group members.  Column 1: detection name; column 2: global profile width measured at 20 per cent of the peak flux density, column 3: inclination as computed using Equation 3, where applicable; column 4: beam-corrected major axis diameter of object in 30 arcsec resolution maps; column 5: heliocentric velocity of the profile midpoint at 20 per cent of the peak flux density; column 6: total dynamical mass, where applicable.  The uncertainty is half the difference between $M_{dyn}$ estimated using $W_{20}$ and the velocity difference across $a_{H_I}$ in Equation 3.  The values for AGC~121467 were not computed due to contaminating foreground star, see text for details.}
\label{dyn}
\begin{tabular}{@{} l c c c c c@{}} 
\hline
Source 		& $W_{20}$ 			&$i$	 &$a_{H_I}$		&$cz_{\odot}$	&$M_{dyn}$ \\
			& (km s$^{-1}$)		&(deg)	&(arcmin)		&(km s$^{-1}$)	&($10^9 M_{\odot}$) \\
(1)			& (2) 				& (3)	&(4)				&(5)				&(6)			\\
\hline
AGC~748849 	& 100 $\pm$ 5 	& 67.8	& 1.7 $\pm$ 0.1	& 3836 $\pm$ 3	& 9 $\pm$ 2\\
NGC~871		& 305 $\pm$ 5 	& 74.4	& 2.2 $\pm$ 0.1	& 3744 $\pm$ 3	& 90 $\pm$ 20\\
AGC~121467	& 32  $\pm$ 5 	&...  	& 1.3 $\pm$ 0.1	& 3776 $\pm$ 3	& ...\\
UGC~1761		& 152 $\pm$ 5 	& 33.7	& 2.2 $\pm$ 0.1	& 3998 $\pm$ 3	& 71 $\pm$ 6\\
AGC~749170 	& 56 $\pm$ 5 	& 63.7	& 2.2 $\pm$ 0.1	& 3905 $\pm$ 3	& 3.2 $\pm$ 0.8\\
AGC~748853	& 109 $\pm$ 5 	& 47.3	& 1.2 $\pm$ 0.1	& 3618 $\pm$ 3	& 12 $\pm$ 3\\
NGC~876		& 407 $\pm$ 5 	& 77.8	& 2.5 $\pm$ 0.1	& 3872 $\pm$ 3	& 180 $\pm$ 6\\
NGC~877		& 436 $\pm$ 5 	& 41.7	& 2.7 $\pm$ 0.2	& 3929 $\pm$ 3	& 500 $\pm$ 10\\
\hline
\end{tabular}
\end{minipage} 
\end{table*}


\section{Stellar Population Analysis}

SExtractor \citep{b1996}, an automated source extraction algorithm, was utilized for the detection and measurement of the optical components to each low {H\sc{i}} mass feature.  Photometric properties for AGC~748849, UGC~1761 and AGC~748853, which all have previously established and well-defined optical counterparts, were readily extracted while processing the entire image.  The optical flux from a foreground star could not be accurately separated from AGC~121467 and consequently, no photometric information has been reported for this source.  The optical properties of the fainter AGC~749170 counterpart were extracted from a 120 $\times$ 110 pixel sub-image centred on this source, with the detection and deblending thresholds tailored to this feature.  The $r'$-band image was used as the reference for data extraction, with the exception of AGC~749170 as its tentative optical counterpart is more apparent in the higher sensitivity $g'$-band image.  We will ultimately use SExtractor to catalog all dwarf galaxy candidate members in the group through statistical source counts; we defer this analysis to a future publication.  Optical spectroscopy, using a similar manner as presented in \citet{d2014}, on each detection in NGC~871/6/7 will also be explored in further follow-up work.

The extracted MegaCam magnitudes for each feature were converted to more conventional SDSS AB magnitudes for comparison to models found in the literature.  The colour differences between each band (i.e. $g-r$ and $g-i$; listed in Table \ref{optical}) were directly compared to \citet{bc2003} stellar population models -- which utilize several stellar evolution prescriptions and various spectral libraries -- for a range of stellar ages and metallicities.  Some objects had varying levels of stellar population degeneracies that could not be mitigated using only three optical bands.  The age and metallicity estimates reported in Table \ref{final} have inherent uncertainties based on the systematics of the \citet{bc2003} models.  The currently available stellar population models likely underestimate the metallicity for dwarf galaxies (see \citealt{h2011a}); nevertheless, the values are sufficient for a relative comparison of metallicity ranges for each low {H\sc{i}} mass group member.  Future optical spectroscopy observations could verify the velocities of the optical sources and provide a more accurate measurement of metallicity values.    

The stellar masses of the associated optical counterparts were computed using the equation:
\begin{equation}
log~M_{*}/[M_{\odot}] = 1.15+ 0.7(g-i) - 0.4 \mbox{M}_i
\end{equation}
where M$_i$ is the absolute magnitude of each object in the $i$-band \citep{t2011}.  Commonly used stellar mass estimation models, such as \citet{b2003}, are quite robust for massive galaxies ($M_* \gtrsim 10^{10} M_{\odot}$) but tend to over-estimate at low masses \citep{h2011a}; however, \citet{t2011} include a census of $10^{7.5}<M_*<10^{8.5} M_{\odot}$ galaxies to calibrate their model, for a better representation of dwarf galaxies.  The computed stellar masses were added to the gas masses to produce total baryonic masses, which -- in comparison to the dynamical mass of each detection -- enables dark matter content estimates.  

Since the $i'$-band component of the faint tentative optical counterpart to AGC~749170 was not detected, an object with the same angular size as that detected in the $g'$-band, assumed to be lying just below the $i'$-band detection threshold, was used to estimate an upper limit $i'$-band magnitude and the associated stellar properties.  For the age and metallicity estimates of AGC~749170, the allowed range for $g-i$ below the upper limit was conservatively chosen to be five times the standard deviation of the $g-i$ values of the other low {H\sc{i}} mass group members (i.e. $g-i$ ranges from 0.7 to 0.4 mag).  The negative value of $g-r$ strongly indicates a younger stellar population for the possible optical component of AGC~749170 and variations in the estimate of $g-i$ have negligible effects on the results. 

\begin{table*}
 \centering
 \begin{minipage}{151mm}
\caption{Optical properties of low {H\sc{i}} mass detections.  AGC~121467 has been omitted from this table due to source extraction inaccuracies from an overlapping foreground star.  Column 1: detection name; column 2: Kron radius containing 50 per cent of the light; column 3: Kron radius containing 90 per cent of the light; column 4: apparent magnitude in $g$-band; column 5: $g-r$ colour; column 6: $g-i$ colour.  The value for AGC~749170 is an upper limit based on the $i$-band detection limit (see text for details about estimations); column 7: absolute magnitude in $i$-band; column 8: stellar mass.}
\label{optical}
\begin{tabular}{@{} l c c c c c c c@{}} 
\hline
Source 			& r$_{50}$	&r$_{90}$	&$g_{SDSS}$			&($g-r$)				&($g-i$)				&M$_{i}$			&$M_{stellar}$\\
				& (arcsec)	&(arcsec)	&(mag)					&(mag)				&(mag)				&(mag)				&($\times 10^8 M_{\odot}$)\\
(1)				& (2) 		& (3)		&(4)					&(5)				&(6)				&(7)				&(8)\\
\hline
AGC~748849	&5.6		&13.2		&16.928 $\pm$ 0.003	&0.359 $\pm$ 0.005	&0.543 $\pm$ 0.006	&-17.1 $\pm$ 0.3	&$2.4\pm0.7$ \\ 
UGC~1761		&17.6		&37.0		&15.300 $\pm$ 0.002	&0.411 $\pm$ 0.002	&0.594 $\pm$ 0.003	&-18.8 $\pm$ 0.4	& $12\pm4$ 	\\
AGC~749170	&5.4 		&7.4		&22.6 $\pm$ 0.4			&-0.5 $\pm$ 0.5		&$<$0.7			&$>$-11				& $<$0.01\\
AGC~748853	&7.7		&17.6		&16.969 $\pm$ 0.004	&0.452 $\pm$ 0.006	&0.659 $\pm$ 0.007	&-17.2 $\pm$ 0.3	& $3.1\pm0.9$\\
\hline
\end{tabular}
\end{minipage} 
\end{table*}

The SFR for each low {H\sc{i}} mass detection was computed using the far-ultraviolet (FUV) flux measured from archival \textit{GALEX} GR6 data -- in the same manner that is presented in Paper 1 -- and the formula from \citet{k1998} with a conversion factor that has been modified for dwarf galaxies with solar ($Z_{\odot}$ = 0.02) or sub-solar metallicities:
\begin{equation}
\mbox{SFR$_{FUV}$[$M_{\odot}$ yr$^{-1}$]} = 1.27 \times 10^{-28} L_{FUV}
\end{equation}
\noindent where $L_{FUV}$ is the luminosity of each object in erg s$^{-1}$ Hz$^{-1}$ \citep{h2010}.  No correction for dust has been made to our SFR estimations.  Although AGC~749170 appears to have a solar to super-solar metallicity, it has no detectable FUV component.  Therefore, its SFR is an upper limit (see Paper 1) and modifications of the conversion factor used have minimal effect on the estimation.  

A final summary of the properties of the gas-rich low {H\sc{i}} mass group members is provided in Table \ref{final}.  The total gas mass for each detection was computed using a factor of 1.33 to account for helium and other elements (see equation 3 from Paper 1).  The average column density $N_{H_I,avg}$ of each detection has been calculated using Equation 1 and the mean intensity measured within the $1 \times 10^{20}$ atoms cm$^{-2}$ contour of the 30 arcsec resolution mom$_0$ maps (see Figs \ref{dwarfs} \& \ref{dark}).

\begin{table*}
 \centering
 \begin{minipage}{176mm}
\caption{Properties of the detected low {H\sc{i}} mass objects in NGC~871/6/7.  Column 1: detection name; column 2: gas mass; column 3: average {H\sc{i}} column density.  The uncertainty is propagated from the standard deviation of the measurements from the 30 arcsec resolution mom$_0$ maps; column 4: baryonic mass; column 5: dynamical to baryonic mass ratio, where applicable; column 6: dynamical mass to $g$-band light ratio, where applicable; column 7: age determined within the systematic uncertainties of the \citet{bc2003} models.  A range of values is presented for objects with degeneracies; column 8: metallicity ($Z_{\odot}$ = 0.02) determined within the systematic uncertainties of the well known models.  Objects with degeneracies are shown with a range of values.  An estimated $g-i$ colour range was used to determine the age and $Z$ values for AGC~749170 (see text for details); column 9: star formation rate for each object.}
\label{final}
\begin{tabular}{@{}l c @{} c @{} c c @{}c c c c@{}} 
\hline
Source 	& $M_{gas}$ 				&$N_{H_I,avg}$				&$M_{baryon}$		&$M_{dyn}/M_{baryon}$ 	&$M/L_g$	&Age			&$Z$		&SFR$_{FUV}$ \\
	& ($\times 10^8 M_{\odot}$)	&($\times10^{20}$ atoms cm$^{-2}$)	&($\times 10^8 M_{\odot}$)	&	&($M_{\odot}/L_{\odot}$)	&($\times10^8$ yrs)	&	&($M_{\odot}$ yr$^{-1}$)\\
(1)				& (2) 				& (3)						&(4)					&(5)				&(6)				&(7)					&(8)				&(9)\\
\hline
AGC~748849 	& 12 $\pm$ 3		& 3.10 $\pm$ 0.07			& 14 $\pm$ 2 		& 6 $\pm$ 2		& 24 $\pm$ 5 	& 8.1				& 0.02			&0.049 $\pm$ 0.007\\
AGC~121467	& 6 $\pm$ 2 		& 3.35 $\pm$ 0.04			& $>$6				&...  			&... 				& ... 				& ...				&0.046 $\pm$ 0.004\\
UGC~1761		& 22 $\pm$ 5		& 5.06 $\pm$ 0.09			& 35 $\pm$ 5		& 21 $\pm$ 4	& 44 $\pm$ 4 	& 32 - 48			& 0.0001		&0.188 $\pm$ 0.007\\
AGC~749170 	& 19 $\pm$ 5		& 2.44 $\pm$ 0.04			& 19 $\pm$ 4		& 1.7 $\pm$ 0.5	 & $>$1000		& 0.063 - 0.12		& 0.02 - 0.05	&$<$0.007\\
AGC~748853	& 10 $\pm$ 3		& 2.72 $\pm$ 0.08			& 14 $\pm$ 2		& 9 $\pm$ 3		& 33 $\pm$ 8  	& 7.2 - 68 	 		& 0.0001 - 0.05	&0.036 $\pm$ 0.004\\
\hline 
\end{tabular}
\end{minipage} 
\end{table*}


\section{Discussion and Conclusions}

We have presented high-resolution {H\sc{i}} observations from the GMRT and deep optical imaging from the CFHT MegaCam of the interacting galaxy group NGC~871/6/7.  Seven of the eight gas-rich detections (three spirals and four dwarfs) have prominent stellar components and appear to be standard dark-matter dominated galaxies that were built hierarchically during the epoch of galaxy assembly.  The other object, AGC~749170, has significant {H\sc{i}} ($M_{H_I} > 10^9 M_{\odot}$) but only a very faint tentative optical counterpart.  Assuming that this gas cloud is self-gravitating and in dynamical equilibrium, then $M_{dyn}/M_{baryon} = 1.7 \pm 0.5$, which indicates a lack of associated dark matter.  An optical feature -- with an apparently young ($\sim$10 Myr) stellar population and a relatively higher metallicity range than the other low {H\sc{i}} mass group members -- that is spatially located within the contour of the peak {H\sc{i}} column density of AGC~749170 has been detected in both the $g'$ and $r'$-band photometry.  These properties corroborate that AGC~749170 is not a primordial dark galaxy and was formed as a result of a previous tidal interaction. 

The middle row of Fig. \ref{dark} shows a clear velocity gradient across AGC~749170; however, this gradient might not be indicative of rotation across the entire source.  In the 15 arcsec resolution images, AGC~749170 breaks up into three ``clumps''.  The smallest northwest clump is likely a tidal extension.  If the two larger southern clumps form a self-gravitating disk, then the optical counterpart should reside between the two lobes rather than in the centre of the southeast clump, where it currently appears.  As well, the second moment maps (bottom row of Fig. \ref{dark}), a measure of the velocity dispersion, would peak near the centre of the two southern clumps.  It is possible that deeper optical observations may reveal a more extensive optical counterpart and that our velocity dispersion maps are noisy.  Nevertheless, our observations suggest that only the southeast clump -- with there is a possible optical counterpart -- is rotating and in dynamical equilibrium while the other {H\sc{i}} two components are tidal extensions.  In this scenario, the self-gravitating feature would have $M_{gas} \approx 8 \times 10^8 M_{\odot}$ and $M_{dyn}/M_{baryon} \sim 1$ as well as a young metal-rich stellar population, which still indicate tidal origins for AGC~749170.

Presuming that a tidal interaction did produce AGC~749170, the global profiles of the large spiral galaxies may provide some insight into its origin.  The fairly symmetrical double-horned profile for NGC~871 suggests that its gas content has had minimal disturbances within the past few Gyr.  Whereas, the {H\sc{i}} components of NGC~876 and NGC~877 overlap both spatially and spectrally while Fig. \ref{GP} shows gas disturbances for each galaxy at heliocentric velocities between 3850 and 4050 km s$^{-1}$.  This range includes the velocity of AGC~749170 ($cz_{\odot} = 3905 \pm 3$ km s$^{-1}$, $W_{20}\approx 60$ km s$^{-1}$).  If NGC~876 and NGC~877, which have a combined {H\sc{i}} mass of $1.5 \times 10^{10} M_{\odot}$, are standard Sc and Sbc spirals galaxies, they would each contain $M_{H_I} \sim 10^{10} M_{\odot}$ \citep{r1994}.  As such, a gaseous cloud of $10^{9} M_{\odot}$ (i.e. AGC~749170) could be added to their combined mass and the two galaxies would still be consistent with field spirals.  Two other dwarf galaxies also appear in the spectral range of the interaction between the two large spirals; however, the projected distance of AGC~748849 ($\sim$250 kpc away) from the current interaction site and the symmetry of the {H\sc{i}} in UGC~1761 (in both the global profile and the gas distribution in relation to its stellar component, see Fig. \ref{dwarfs}c) suggest that these dwarfs are neither involved in the current interaction nor are they likely progenitors of AGC~749170.

Based on the dynamical masses of the suspected parent galaxies (NGC~876 and NGC 877), AGC~749170 is likely the result of a 2.5:1 mass ratio interaction event.  Major mergers in both observations and numerical simulations have been shown to produce mass condensations of $10^{9} M_{\odot}$ located at the tips of extended tidal tails ($\sim$10$-$100 kpc in length; e.g.~\citealt{b2004}, \citealt{m2011}).  AGC~749170 is located at a projected distance of $\sim$100 kpc from its possible progenitors; however, the lack of a tidal tail and a readily detectable stellar component precludes its classification as a typical TDG or tidal knot.  

The systemic velocity difference between AGC~749170 and the two large spirals is sufficient enough to have possibly moved the cloud to its current separation distance over the past few Gyr, during which time any gaseous trace of a tidal tail would have faded away (\citealt{hol2011}).  Nevertheless, if a cloud of $M_{H_I} >10^{9} M_{\odot}$ has existed for such a length of time, its low stellar content and young stellar age ($M_{stellar} \lesssim 10^{6} M_{\odot}$, $t \sim 9$ Myr) is unusual compared to other tidal objects with similar gas masses (e.g.~\citealt{s2004}, \citealt{b2007}).  It is also valid to note that reflection halo in the optical photometry due to HD 14192, a bright foreground star, could be obscuring a possible stellar tidal tail in that region (see Fig.~\ref{CFHT}). 

The critical column density below which no star formation is expected to occur is $N_{H_I,crit} = (3-10) \times10^{20}$ atoms cm$^{-2}$ \citep{s2004}.  As reported in Table \ref{final}, AGC~749170 has an average column density of $(2.44 \pm 0.04) \times10^{20}$ atoms cm$^{-2}$ within the lowest contour of the 30 arcsec resolution {H\sc{i}} maps (which falls below the critical threshold) but at 15 arcsec resolution, $N_{H_I,avg} = (4.6 \pm 0.1) \times10^{20}$ atoms cm$^{-2}$ and its peak column density, which coincides with the spatial location of its possible optical counterpart, is well above this value.  In a recent study of stellar formation within tidal debris, \citet{may2007} found that two tidal tails with $N_{H_I} \geq 4 \times10^{20}$ atoms cm$^{-2}$ have no significant star cluster populations.  In that paper, the gas volume density is presented as a more useful diagnostic tool, where one of the tails in NGC 4038/9 has a high column density that spans a small area ($\sim$10 kpc$^2$), which makes its unlikely to form a significant amount of stars.  The high column density region of AGC~749170 spans at least 300 kpc$^{2}$ and even though our deep optical imaging clearly shows star clusters for each of the other low {H\sc{i}} mass group members, AGC~749170 remains extremely optically dim.

Table \ref{otherdark} compares various dark gas-rich galaxy-like features from the literature.  It appears that AGC~749170 is not unique and shares similar properties to other puzzling objects, in particular the SW clump in {H\sc{i}} 1225+01 (assumed to be tidal by \citet{c1995}, but recently revisited as being a possible dark galaxy by \citet{m2012}) and Vela C \citep{e2010}.  The three clouds have roughly the same {H\sc{i}} mass, velocity width and separation distance from their likely parent galaxies.  AGC~749170 and Vela C also have comparable diameters and a similar ``clumpliness'' in their {H\sc{i}} maps.  The lower {H\sc{i}} column density of Vela C could be attributed to smearing effects from the larger beam size, as implied by Equation 1.  Considering the fact that the tentative optical counterpart to AGC~749170 is barely visible in deep CFHT photometry, similar sensitivity observations for Vela C are required to search for a comparable optical feature.  The SW clump of {H\sc{i}} 1225+01 appears to have an {H\sc{i}} bridge linking it to the NE clump, the latter shows clear signs of tidal interactions \citep{c1995_2}.  Nevertheless, deep optical imaging (down to a limit of $R_{AB}$ = 28.3 mag arcsec$^{-2}$) of the region has not revealed any optical component to the SW clump \citep{m2012}.  Overall, the star formation properties of these clouds are extreme compared to typical tidal dwarfs but this result could be the effect of the group environment.  

\begin{table*}
 \centering
 \begin{minipage}{197mm}
\caption{Properties of various ``dark'' gas-rich galaxy-like features.  Currently, AGC~749170 is the only object in this list that has a tentative optical counterpart.  Column 1: object name; column 2: associated galaxy group/low density cluster region, {H\sc{i}} system or name of nearest neighbour; column 3: distance of object from the sun as reported in each paper.  Note, the distance to GBT 1355+5439 is currently unknown, we use the values presented under the assumption that this object is tidal a remnant near M101; column 4: major axis FWHM of the synthesized beam for {H\sc{i}} observations in kpc based on the distance of the object; column 5: {H\sc{i}} mass; column 6: peak {H\sc{i}} column density.  For values not reported in the literature $N_{H_I,peak}$ is assumed to be greater than or equal to the highest {H\sc{i}} contour in the $mom_0$ maps; column 7: velocity width at half the peak flux density.  For values not reported in the literature $W_{50}$ is assumed to be less than the velocity range of the object in the $mom_1$ maps; column 8: major axis diameter of object; column 9: projected linear distance to nearest neighbour of possible tidal interaction; column 10: detection limit of optical photometry in various bands where (aside from AGC 749170) no optical counterpart has been detected; column 11: references abbreviated as: C95=\citealt{c1995_2}, W05=\citealt{w2005_2}, E10=\citealt{e2010}, K10=\citealt{k2010}, M12=\citealt{m2012}, O13=\citealt{o2013_2}, S13=\citealt{s2013}. \label{otherdark}}
\label{otherdark}
\begin{tabular}{@{}l@{} c c c@{} c@{} c@{} c@{} c@{} c@{} c@{} c @{}} 
\hline
Name				&Group 	&d		&$\theta_{H_I,maj}$	& $M_{H_I}$ 		&$N_{H_I,peak}$	&$W_{50}$	&$a_{H_I}$		&d$_{NN}$	 &Image depth		&Reference\\
					&		&(Mpc)	&(kpc)			& ($\times 10^8 M_{\odot}$)	&($\times10^{20}$ atoms cm$^{-2}$)	&(km s$^{-2}$)	&(kpc)	&(kpc)		& (mag arcsec$^{-2}$)	&\\
(1)					& (2) 	& (3)		&(4)				&(5)				&(6)				&(7)				&(8)				&(9)				&(10)				&(11)\\
\hline
AGC 749170 			& NGC 871/6/7		&50		&3.7		&14 $\pm$ 4 		& 13 $\pm$ 1		& 46 $\pm$ 6 		& 32 $\pm$ 2 		& $\sim$90		&CFHT $g' = 27.3$	& this work\\
SW clump			&{H\sc{i}} 1225+01	&20		&3.9		&8.6				&11				&34 $\pm$ 7		& $<100$			& $\sim$100		&MOA $R=28.3$		& C95 \& M12\\	
HIJASS J1021+6842		&M81			&4		&1.17	&1.5				&$\sim$1.8		&$\sim$120		& $>30$			&110				&DSS $B \sim24$		& W05\\
Vela A				&NGC 3256/63		&38		&10.7	&1.6 $\pm$ 0.5 	& 0.3 $\pm$ 0.1 	& 41 $\pm$ 6		& 26 $\pm$ 3		& $\sim$80		&DSS $B \sim24$ 		& E10\\		
Vela B				&NGC 3256/63		&38		&10.7	& 14 $\pm$ 5		& 1.2 $\pm$ 0.5 	& 100 $\pm$ 10	& 55 $\pm$ 4		& $\sim$80		&DSS $B \sim24$		& E10\\		
Vela C				&NGC 3256/63		&38		&10.7	& 16 $\pm$ 5 		& 1.7 $\pm$ 0.7	& 44 $\pm$ 1		& 48 $\pm$ 2		& $\sim$130		&DSS $B \sim24$		& E10\\		
C1 North				&Virgo Periphery	&16.7	&2.0		& 0.19 $\pm$ 0.02	& $\geq 3$		& 22 $\pm$ 6		& 5 $\pm$ 1		& 276			&SDSS $g\sim26$		& K10\\	
C1 South				&Virgo Periphery	&16.7	&2.0		& 0.25 $\pm$ 0.02	& $\geq 2$		& 20 $\pm$ 8		& 7 $\pm$ 1		& 276			&SDSS $g\sim26$ 		& K10\\	
C2 North				&Virgo Periphery	&34.8	&4.2		& 0.40 $\pm$ 0.06	& $\geq 1.2$		& 13 $\pm$ 4		& 25.3 $\pm$ 0.4	& 40				&SDSS $g\sim26$		& K10\\	
C2 West				&Virgo Periphery	&34.8	&4.2		& 0.71 $\pm$ 0.06	& $\geq 1.3$		& 41 $\pm$ 9		& 24.3 $\pm$ 0.3	& 40				&SDSS $g\sim26$		& K10\\	
C2 South				&Virgo Periphery	&34.8	&4.2		& 0.14 $\pm$ 0.03	& $\geq 1.1$		& 6 $\pm$ 5		& 8.1 $\pm$ 0.2	& 40				&SDSS $g\sim26$		& K10\\	
GBT 1355+5439		&M101			&6.9		&1.8		& 0.11			& 0.71			& $<15$			& $\sim$10		& 150			&Burrell Schmidt $V$ = 29	&O13\\	
$C_{\mbox{S}}$		&HCG 44			&25		&6.4		& 1.2				&$\geq 0.9$		& $<100$			&$\sim$30		& $\sim$50		&CFHT $g'$ $\sim28.5$	&S13\\	
\hline 
\end{tabular}
\end{minipage} 
\end{table*}

AGC~749170, the SW clump of {H\sc{i}} 1225+01, Vela B and Vela C appear to have relatively higher gas masses than other compact (i.e. not diffuse and/or extended structures as found in \citealt{r2001}) and optically dark clouds.  It is possible that these objects lie on the extreme end of a rare class of dark/dim tidal features.  Hydrodynamic simulations could be used to determine the likelihood of a $10^{9} M_{\odot}$ gas cloud moving $\sim$100 kpc from its location of origin while dissipating the tidal tail and remaining relatively star-less.  A future paper could also investigate the interplay between a merging event, the intrinsic group dynamics or the overall tidal effect of the environment in producing such objects as most dark/dim features are detected in groups or on the outskirts of clusters rather than in isolation. 


\vspace{3mm}
We thank the staff of the GMRT that made our interferometric observations possible.  Thank-you to the reviewer for his/her suggestions to improve the clarity of this paper.  We also thank J.~A. Irwin for her input on the overall research project.
K.~S. acknowledges funding from the National Sciences and Engineering Research Council of Canada.   The ALFALFA team at Cornell is supported by U.S. NSF grants  AST-0607007 and AST-1107390 to R.~G. and M.~P.~H. and by grants from the Brinson Foundation. The GMRT is run by the National Centre for Radio Astrophysics of the Tata Institute of Fundamental Research.  This research used optical observations obtained with MegaPrime/MegaCam, a joint project of CFHT and CEA/DAPNIA, at the Canada-France-Hawaii Telescope (CFHT) which is operated by the National Research Council (NRC) of Canada, the Institute National des Science de l'Univers of the Centre National de la Recherche Scientifique of France, and the University of Hawaii.

\end{document}